# What in fact proves the violation of the Bell-type inequalities?

*by Sofia Wechsler*

**Abstract**
A. Peres constructed an example of particles entangled in the state of spin singlet. He claimed to have obtained the CHSH inequality and concluded that the violation of this inequality shows that in a measurement in which some variables are tested, other variables, not tested, have no defined value.
In the present paper is proved that the correct conclusion of the violation of the CHSH inequality is different. It is proved that the classical calculus of probabilities of test results, obeying the Kolmogorov axioms, is unfit for the quantum formalism, dominated by probability amplitudes.

## 1. Introduction

J. S. Bell developed in his famous work [1] an inequality for three observables, and it was extended to more observables [2]. These inequalities were found to disagree with the quantum mechanics (QM). The violation of these inequalities was proved by the experiment, [3] – [11], and is typically interpreted as indicating that the quantum world allows nonlocal influences between measurement results of entangled quantum objects [1], [3] - [6], [12] - [18] (see also the references in [18] and also a generalization of the Bell-type inequalities in [19] ).

In his famous article "Unperformed experiments have no results" [20] (see also [21]) A. Peres claimed to have deduced from the violation of Bell-type inequalities a new conclusion, namely, that incompatible observables cannot take definite values in the same test. By incompatible is understood that the operators corresponding to the observables don't commute. So, Peres concluded that if the observables *A* and *B* are measured and produce definite values, *the wave-function* (w-f) *cannot predict* definite values for non-measured observables, *A'* and *B'*, not commuting with *A* respectively *B*.

It may though be that there are effects beyond the predictions of the w-f, i.e. that the QM admits a substructure. For example, if the quantum object behaves according to Bohm's mechanics, [22], some observables may take values although the w-f does not acknowledge them.

It is argued in the present article that what in fact is proved by the violation of the Bell-type inequalities is another feature, namely, that the classical calculus of distribution of probabilities, satisfying the Kolmogorov axioms, is incompatible with the predictions of the w-f. The assumption of locality of measurement results yields just a particular case of such a distribution.

The next sections are organized as follows: section 2 defines the most general distribution of probabilities of variables' values, for which the Bell-type inequalities can be derived; no locality assumption is made. Section 3 derives the CHSH inequality using this general distribution. Section 4 compares the consequences of this distribution with the quantum predictions, and a contradiction is found. Section 5 contains conclusions.

## 2. A probability distribution under the Komogorov axioms

Consider a set of events $\{E_1, E_2, \ldots, E_n\}$ with probabilities $P$ from a common space satisfying the Kolmogorov axioms

1. $0 \leq P(E_i) \leq 1$,
2. $P(E_i \text{ OR } E_j \text{ OR} \ldots \text{OR } E_k) = P(E_i) + P(E_j) + \ldots + P(E_k)$.

In this text we will have to do with four binary variables: $A$ and $A'$ belong to a photon 1, $B$ and $B'$ belong to a photon 2. All the four variables are polarizations, $A$ and $A'$ along the directions **a** respectively **a'**, $B$ and $B'$ along the directions **b** respectively **b'**. We will denote the polarization along a direction **u** by "$+_u$" and perpendicular to **u** by "$\neg_u$". So, in a measurement of the variable $A$ for photon 1 and $B$ for photon 2 we will get the result "$+_a$" for $A$ and "$\neg_b$" for $B$ with the probability $P(+_a \neg_b)$.

*Assumption A*: in a test of the two particles, each one along some direction, all the four variables take definite values, i.e. the non-measured variables take values as well as the measured ones.

Therefore we will have to do with probabilities of the form $P_{AB'}(+_a +_{a'} +_b -_{b'})$ where the indexes $A$ and $B'$ after $P$ indicate which observables were in fact measured.
Applying then the Komogorov axioms one has

$$P_{AB'}(+_a -_{b'}) = P_{AB'}(+++-) + P_{AB'}(++--) + P_{AB'}(+-+-) + P_{AB'}(+---). \quad (1)$$

We omitted here and so we will do in the rest of the article, the indexes inside the round parentheses in the four-variable probabilities, but we will keep the convention that the order of the results is $\pm_a \pm_{a'} \pm_b \pm_{b'}$.

*Assumption B*: the distribution of four-variable probabilities is independent of which variables are actually measured.

That means

$$P_{AB}(q_a\, q_{a'}\, q_b\, q_{b'}) = P_{A'B}(q_a\, q_{a'}\, q_b\, q_{b'}) = P_{AB'}(q_a\, q_{a'}\, q_b\, q_{b'}) = P_{A'B'}(q_a\, q_{a'}\, q_b\, q_{b'}), \quad (2)$$

with $q_a, q_{a'}, q_b, q_{b'} \in [+,-]$. That gives us the possibility to omit the specification of the actual measurement in the four-variable probability, e.g. we will write simply $P(q_a\, q_{a'}\, q_b\, q_{b'})$. We will also omit the specification of the actual measurement in the two-variable measurement, because it appears inside the parentheses, i.e. $P_{AB}(q_a\, q_b) = P(q_a\, q_b)$.

## 3. A general derivation of the CHSH inequality

With the distribution of probability developed in the previous section we can obtain the following results

$$P(+_a +_b) = P(+++ +) + P(+++ -) + P(+-++) + P(+-+ -),$$
$$P(+_a -_b) = P(++-+) + P(++--) + P(+--+) + P(+---),$$
$$P(-_a +_b) = P(-+++) + P(-++-) + P(--++) + P(--+-),$$
$$P(-_a -_b) = P(-+-+) + P(-+--) + P(---+) + P(----),$$

The average $\overline{AB}$ of the results of the joint measurement of $A$ and $B$ can be expressed according to the above equalities as

$$\begin{aligned}\overline{AB} = &\ \{P(++++) + P(+++-) + P(+-++) + P(+-+-)\} \\ &- \{P(++-+) + P(++--) + P(+--+) + P(+---)\} \\ &- \{P(-+++) + P(-++-) + P(--++) + P(--+-)\} \\ &+ \{P(-+-+) + P(-+--) + P(---+) + P(----)\}\end{aligned} \quad (3)$$

One can similarly express in terms of four-variable probabilities, the averages $\overline{A'B}$, $\overline{AB'}$, $\overline{A'B'}$

$$\begin{aligned}\overline{A'B} &= P(+_{a'} +_b) - P(+_{a'} -_b) - P(-_{a'} +_b) + P(-_{a'} -_b), \\ \overline{AB'} &= P(+_a +_{b'}) - P(+_a -_{b'}) - P(-_a +_{b'}) + P(-_a -_{b'}), \\ \overline{A'B'} &= P(+_{a'} +_{b'}) - P(+_{a'} -_{b'}) - P(-_{a'} +_{b'}) + P(-_{a'} -_{b'}).\end{aligned} \quad (4)$$

e.g.

$$\begin{aligned}\overline{A'B} = &\ \{P(++++) + P(+++-) + P(-+++) + P(-++-)\} \\ &- \{P(++-+) + P(++--) + P(-+-+) + P(-+--)\} \\ &- \{P(+-++) + P(+-+-) + P(--++) + P(--+-)\} \\ &+ \{P(+--+) + P(+---) + P(---+) + P(----)\}.\end{aligned} \quad (5)$$

Since $\sum_{A, A', B, B'} P(A, A', B, B') = 1$, one has

$$\begin{aligned}&P(++++) + P(+++-) + P(++-+) + P(++--) \\ &+ P(+-++) + P(+-+-) + P(+--+) + P(+---) \\ &+ P(-+++) + P(-++-) + P(-+-+) + P(-+--) \\ &+ P(--++) + P(--+-) + P(---+) + P(----) = 1\end{aligned} \quad (6)$$

From the equalities (3) and (4) one can find by direct substitution and reducing identical terms that

$$\begin{aligned}\overline{AB}+\overline{A'B}-\overline{AB'}+\overline{A'B'} = 2\{&P(+++ +)+P(+++-)-P(++-+)-P(++--)\\&-P(+-++)+P(+-+-)-P(+--+)+P(+---)\\&+P(-+++)-P(-++-)+P(-+-+)-P(-+--)\\&-P(--++)-P(--+-)+P(---+)+P(----)\}.\end{aligned} \quad (7)$$

Comparing with (6) one can see that

$$\overline{AB}+\overline{A'B}-\overline{AB'}+\overline{A'B'} \le 2 . \quad (8)$$

Indeed, if the sum of all the four-variable probabilities is equal to 1, the sum in which some of the probabilities appear with minus should be in general less than 1.
The inequality (8) is the CHSH inequality.

## 4. Then what in fact prove the Bell-type inequalities?

It is known that the CHSH inequality is violated if one considers the photon singlet. For a measurement of the polarization of the photon 1 along a direction **u**, and a measurement of the polarization of photon 2 along a direction **v**, the averaged product of the two results is equal to $\cos(2\theta_{uv})$, where $\theta_{uv}$ is the angle between the two directions. Choosing for the observables $A$ and $A'$ the polarization of the photon 1 along the directions **a** and **a'**, and for the observables $B$ and $B'$ the polarization of the photon 2 along the directions **b** and **b'**, as in the figures 1 and 2, the quantum formalism predicts

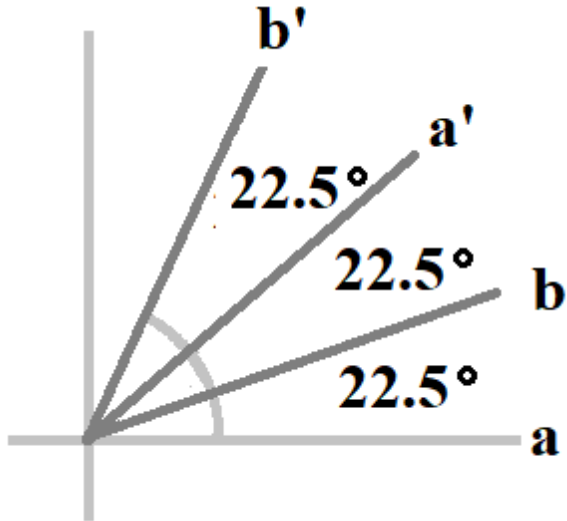

Figure 1. Four coplanar directions.
See explanations in the text.

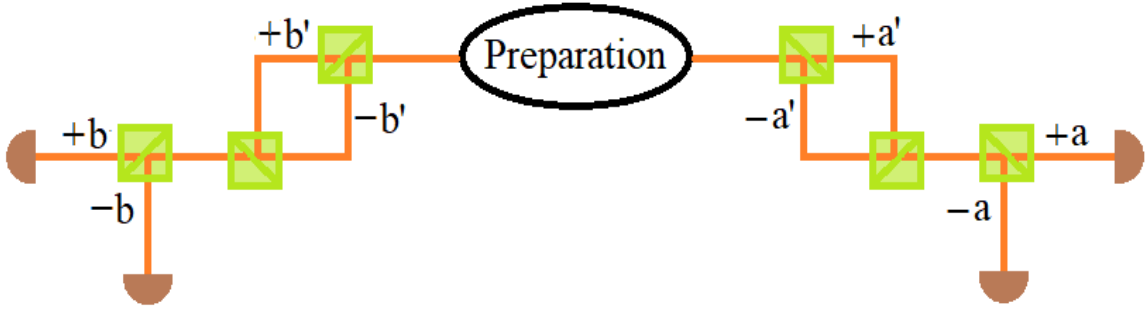

Figure 2. Arrangement of a CHSH experiment involving different polarization directions. The wave-packet of the photon 1 (2) is split according to the polarization directions **a'** and perpendicular to **a'** (**b'** and perpendicular to **b'**). Then, the two wave-packets are merged back into one and split according to the polarization axes **a** and perpendicular to **a** (**b** and perpendicular to **b**).

$$\langle AB \rangle + \langle A'B \rangle - \langle AB' \rangle + \langle A'B' \rangle = 2\sqrt{2}, \tag{9}$$

This result disagrees with the inequality (8).

This violation proves that some of the assumptions made in the derivation of the distribution $P(q_a\ q_{a'}\ q_b\ q_{b'})$ in section 2 must be wrong. These assumptions are the Kolmogorov axioms 1 and 2 and the assumptions A and B. Some of them must disagree with the QM, and it is known that QM was never disproved by the experiment.

Whether the assumptions A and B disagree with the QM is not obvious. But it is obvious that the axiom 2 of Kolmogorov does disagree with the QM. Indeed, by this axiom the probability of occurrence of a result $q_1$ or $q_2$ satisfies

$$\text{Prob}(q_1 \text{ OR } q_2) = \text{Prob}(q_1) + \text{Prob}(q_2). \tag{10}$$

However, in the quantum formalism the basic element in the calculus of probability of a result $q$ is not the probability Prob($q$), but the amplitude of probability $\boldsymbol{a}(q)$, with

$$\text{Prob}(q_1) = |\boldsymbol{a}(q_1)|^2,\ \text{Prob}(q_2) = |\boldsymbol{a}(q_2)|^2, \tag{11}$$

and instead of (10) one has by the QM

$$\text{Prob}(q_1 \text{ OR } q_2) = |\boldsymbol{a}(q_1) + \boldsymbol{a}(q_2)|^2 = \text{Prob}(q_1) + \text{Prob}(q_2) + \boldsymbol{a}(q_1)^*\boldsymbol{a}(q_2) + \boldsymbol{a}(q_2)^*\boldsymbol{a}(q_1). \tag{12}$$

Let's exemplify the problem on the probability $P(+_a +_b)$, i.e. the actual measurement is done in the base $\{+_a,-_a\}$ for photon 1 and $\{+_b,-_b\}$ for photon 2, and we retain the pairs producing the response "++". We will show that QM and the Kolmogorov-type calculus yield different expressions, as do (10) and (12). For this purpose we write the w-f of the photon singlet putting in evidence the polarizations along **a'**, **a**, and **b'**, **b**, and we denote by $\theta_{uv}$ the angle between directions **u** and **v**.

So, we start from the expression of the w-f in the polarization base $\{+_{a'},-_{a'}\}$ for photon 1, and $\{+_{b'},-_{b'}\}$ for photon 2

$$|\psi\rangle = \frac{1}{\sqrt{2}}\{\cos(\theta_{a'b'})|+_{a'}\rangle|+_{b'}\rangle + \sin(\theta_{a'b'})|+_{a'}\rangle|-_{b'}\rangle$$
$$-\sin(\theta_{a'b'})|-_{a'}\rangle|+_{b'}\rangle + \cos(\theta_{a'b'})|-_{a'}\rangle|-_{b'}\rangle\} \quad (13)$$
$$= a(+_{a'}+_{b'})|+_{a'}\rangle|+_{b'}\rangle + a(+_{a'}-_{b'})|+_{a'}\rangle|-_{b'}\rangle$$
$$+ a(-_{a'}+_{b'})|-_{a'}\rangle|+_{b'}\rangle + a(-_{a'}-_{b'})|-_{a'}\rangle|-_{b'}\rangle.$$

Following the evolution of the pair in the apparatus – see figure 2 – let's pass to the base $\{+_a,-_a\}$ for photon 1

$$|+_{a'}\rangle = \cos(\theta_{a'a})|+_a\rangle + \sin(\theta_{a'a})|-_a\rangle$$
$$|-_{a'}\rangle = -\sin(\theta_{a'a})|+_a\rangle + \cos(\theta_{a'a})|-_a\rangle \quad (14)$$

$$|\psi\rangle = \frac{1}{\sqrt{2}}\{\cos(\theta_{a'b'})[\cos(\theta_{a'a})|+_a\rangle|+_{b'}\rangle + \sin(\theta_{a'a})|-_a\rangle|+_{b'}\rangle]$$
$$+ \sin(\theta_{a'b'})[\cos(\theta_{a'a})|+_a\rangle|-_{b'}\rangle + \sin(\theta_{a'a})|-_a\rangle|-_{b'}\rangle]$$
$$- \sin(\theta_{a'b'})[-\sin(\theta_{a'a})|+_a\rangle|+_{b'}\rangle + \cos(\theta_{a'a})|-_a\rangle|+_{b'}\rangle]$$
$$+ \cos(\theta_{a'b'})[-\sin(\theta_{a'a})|+_a\rangle|-_{b'}\rangle + \cos(\theta_{a'a})|-_a\rangle|-_{b'}\rangle]\} \quad (15)$$
$$= a(+_{a'}+_a+_{b'})|+_a\rangle|+_{b'}\rangle + a(+_{a'}-_a+_{b'})|-_a\rangle|+_{b'}\rangle$$
$$+ a(+_{a'}+_a-_{b'})|+_a\rangle|-_{b'}\rangle + a(+_{a'}-_a-_{b'})|-_a\rangle|-_{b'}\rangle$$
$$+ a(-_{a'}+_a+_{b'})|+_a\rangle|+_{b'}\rangle + a(-_{a'}-_a+_{b'})|-_a\rangle|+_{b'}\rangle$$
$$+ a(-_{a'}+_a-_{b'})|+_a\rangle|-_{b'}\rangle + a(-_{a'}-_a-_{b'})|-_a\rangle|-_{b'}\rangle.$$

Finally we pass from the base $\{+_{b'},-_{b'}\}$ for the photon 2 to the base $\{+_b,-_b\}$

$$|+_{b'}\rangle = \cos(\theta_{b'b})|+_b\rangle + \sin(\theta_{b'b})|-_b\rangle$$
$$|-_{b'}\rangle = -\sin(\theta_{b'b})|+_b\rangle + \cos(\theta_{b'b})|-_b\rangle, \quad (16)$$

and get

$$|\psi\rangle = \{a(+\,+\,+\,+) + a(+\,+\,+\,-) + a(+\,-\,+\,+) + a(+\,-\,+\,-)\}|+_a\rangle|+_b\rangle + \ldots \quad (17)$$

where the order inside the round parentheses is $q_a, q_{a'}, q_b, q_{b'}$, and

$$a(+++ +) = \cos(\theta_{a'b'})\cos(\theta_{a'a})\cos(\theta_{b'b})/\sqrt{2}$$
$$a(+++ -) = -\sin(\theta_{a'b'})\cos(\theta_{a'a})\sin(\theta_{b'b})/\sqrt{2}$$
$$a(+-++) = \sin(\theta_{a'b'})\sin(\theta_{a'a})\cos(\theta_{b'b})/\sqrt{2}$$
$$a(+-+-) = \cos(\theta_{a'b'})\sin(\theta_{a'a})\sin(\theta_{b'b})/\sqrt{2}$$

(18)

According to the QM from these amplitudes of probability one can calculate the probability of getting the result $+_a +_b$

$$P(+_a +_b) = |a(+++ +) + a(+++ -) + a(+-++) + a(+-+-)|^2. \quad (19)$$

Substituting in (19) the expressions in (18) one gets after a simple calculus of trigonometry

$$P(+_a +_b) = \cos^2(\theta_{ab})/2. \quad (20)$$

However, if we equate $P(q_a\ q_{a'}\ q_b\ q_{b'}) = |a(q_a\ q_{a'}\ q_b\ q_{b'})|^2$, one can immediately see that the sum of the absolute squares of the amplitudes in (18) does not lead to the quantum result (20).

A similar problem of disagreement between the classical probability calculus and the quantum formalism was posed by Tarozzi [23] and later by Božić et al. [24]. Both referred to the two-slit experiment in which the intensity of the image on a screen $\mathcal{S}$ beyond the slits $B$ and $C$, $|\psi(\mathbf{r},t)|^2$, differs from the sum of the intensities of the images created with only one of the slits open, $|\varphi_B(\mathbf{r},t)|^2$ and $|\varphi_C(\mathbf{r},t)|^2$. Indeed, $|\psi(\mathbf{r},t)|^2 = |\varphi_B(\mathbf{r},t) + \varphi_C(\mathbf{r},t)|^2 = |\varphi_B(\mathbf{r},t)|^2 + |\varphi_C(\mathbf{r},t)|^2 + X(\mathbf{r},t)$, where $X(\mathbf{r},t)$ is the interference term $X(\mathbf{r},t) = 2\operatorname{Re}[\varphi_B^*(\mathbf{r},t)\varphi_C(\mathbf{r},t)]$.

For solving the dilemma why the whole differs from the sum of its parts, Božić et al. suggested to regard the experiment in which both sits are open as having a different configuration than the experiments in which only one slit is open and in consequence, what passes through the apparatus is different in the two types of experiment. They suggested to write for the interference experiment $|\psi(\mathbf{r},t)|^2 = |\tilde{\varphi}_B(\mathbf{r},t)|^2 + |\tilde{\varphi}_C(\mathbf{r},t)|^2$, where each one of $\tilde{\varphi}_B(\mathbf{r},t)$ and $\tilde{\varphi}_C(\mathbf{r},t)$ contains also a part from $X(\mathbf{r},t)$.

This idea seems appealing for the experiment in the figure 2 too. That means, to construct a distribution of probabilities in which $P(q_a\ q_{a'}\ q_b\ q_{b'})$ contains besides the absolute square of the respective four-variable amplitude, a part of the interference terms.

But that turns to be impossible. Let's take as an example the probability $P(q_a\ q_{a'}\ q_b\ q_{b'}) = |+-++\rangle$, which according to (18) should be equal to $\sin^2(\theta_{a'b'})\sin^2(\theta_{a'a})\cos^2(\theta_{b'b})$. This expression cannot be modified since it is dictated by the Malus law. It says that if the particle 1 took the path –a', the particle 2 takes the path +b' with probability $\sin^2(\theta_{a'b'})$. Next, the particle 1 takes the path +a with probability

$\sin^2(\theta_{a'a})$ and the particle 2 takes the particle +b with probability $\cos^2(\theta_{b'b})$. There is no possibility for changing $P(q_a\, q_{a'}\, q_b\, q_{b'})$.

## Conclusions

In his article "Unperformed experiments have no results", [20], Peres concluded that the assumption A – see section 2 – is wrong. However, as shown here, the correct cause of the violation of the CHSH inequality is the incompatibility between the classical calculus of probabilities, governed by the Kolmogorov axioms, and the quantum probabilities governed by the calculus of amplitudes of probability.

It has to be mentioned that 16 years after the publication of Peres' article, Berndl et Goldstein also concluded, [25], in an examination of the celebrated Hardy's article "Quantum Mechanics, Local Realistic Theories, and Lorenz-Invariant Realistic Theories", [26], that observables that were not measured may have no definite values in a test in which other observables were actually measured. But, again, this conclusion does not result from the violation of the Bell-type inequality.

No locality was assumed in the derivation of the distribution $P(q_a\, q_{a'}\, q_b\, q_{b'})$ obtained here, The hypothesis of locality of measurement results leads to a particular case of $P(q_a\, q_{a'}\, q_b\, q_{b'})$. The distribution derived here covers not only distributions obtained in base of the locality assumption, but also distributions of results dependent both on a local setup as well as on a distant setup.

The fact that the quantum probabilistic formalism is based on superposition of amplitudes of probability stresses the wave nature of the quantum object.